\documentclass[letter]{aa}
\usepackage{natbib}
\usepackage{graphicx}
\usepackage{txfonts}

\newcommand{\be}{\begin{equation}}
\newcommand{\ee}{\end{equation}}
\newcommand{\bea}{\begin{eqnarray}}
\newcommand{\eea}{\end{eqnarray}}  
\newcommand{\D}{{\rm d}}
\newcommand{\E}{{\rm e}}

\newcommand{\BM}{{\rm B}}
\newcommand{\RC}{{\rm R}}

\newcommand{\zmena}[1]{#1}

\begin{document}

\title{Hydrogen Balmer line formation in solar flares affected\\
by return currents}
\titlerunning{Balmer line and return current}

\author
{
J.\,\v{S}t\v{e}p\'an\inst{1}\fnmsep\inst{2}
\and J.\,Ka\v{s}parov\'a\inst{1}
\and M.\,Karlick\'y\inst{1}
\and P.\,Heinzel\inst{1}
}

\institute
{
Astronomical Institute, Academy of Sciences of the Czech Republic, v.v.i.,
Fri\v{c}ova 298, 251\,65 Ond\v{r}ejov, Czech Republic\\
\email{[stepan;kasparov;karlicky;pheinzel]@asu.cas.cz}
\and LERMA, Observatoire de Paris -- Meudon, CNRS UMR 8112, 5,
place Jules Janssen, 92195 Meudon Cedex, France\\
\email{jiri.stepan@obspm.fr}
}

\date{Received 27 June 2007 / Accepted 30 July 2007}

\abstract
{}
{
We investigate the effect of the electric return currents in solar flares
on the profiles of hydrogen Balmer lines. We consider the monoenergetic
approximation for the primary beam and runaway model
of the neutralizing return current.
}
{
Propagation of the 10~keV electron beam from a coronal reconnection site
is considered for the semiempirical chromosphere model F1.
We estimate the local number
density of return current using two approximations for beam energy
fluxes between $4\times 10^{11}$ and
$1\times 10^{12}\,{\rm erg\,cm^{-2}\,s^{-1}}$.
Inelastic collisions of
beam and return-current electrons with hydrogen are included according to
their energy distributions, and the hydrogen Balmer line intensities are
computed using an NLTE radiative transfer approach.
}
{
In comparison to traditional NLTE models of solar flares that neglect
the return-current effects, we found a significant increase emission
in the Balmer line cores due to nonthermal excitation by return current.
Contrary to the model without return current, the line shapes are
sensitive to a beam flux. It is the result of variation in the
return-current energy that is close to the hydrogen excitation thresholds and
the density of return-current electrons.
}
{}
       

\maketitle

\section{Introduction}

The ongoing study of nonthermal excitation of the flaring chromospheric plasmas
has been mainly concentrated on the effect of particle beams coming from
the coronal reconnection site
\citep[][and references therein]{canfield84,hawley94,fang93,kasparova02,
stepan07protons}.
However, \citet{karlicky02} and
\citet{karlicky04} recently suggested that the role of neutralizing return currents can be
as important as the role of the primary beam itself, both for intensity and
linear polarization profiles.
\citet{karlicky04} proposed a simple model of return current formed by the
runaway electrons and compared the rates of atomic transitions due to
collisions both with the thermal electrons and with the electrons of
the primary beam,
and due to collisions with the return current formed by the runaway electrons.
They showed that the rates due to the return current would dominate the
collisional processes in the atmospheric region of Balmer line formation.
However, no calculations of theoretical spectral line profiles were presented.

The aim of this paper is to take a first step towards
self-consistent modeling of the Balmer line formation
with return-current effects taken into account.
We use a semi-empirical model of
the flaring atmosphere as a basis for our NLTE radiative transfer model.
Then we use a standard model for electron-beam deceleration due to Coulomb
collisions with the ambient 
atmosphere and combine it with the two different physical
models of the return-current generation. We incorporate the relevant processes
that enter the atomic statistical equilibrium equations and solve them
with the non-local equations of radiation transfer.
At the end, we discuss the results and validity of our models.

\begin{figure}[!ht]
\centering
\includegraphics[width=\columnwidth]{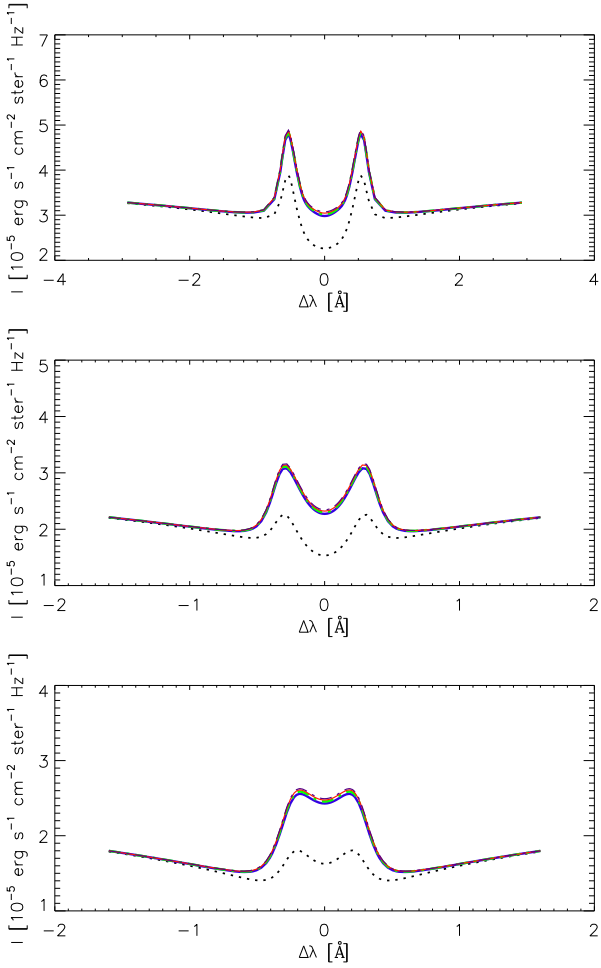}
\caption{
From upper panel: H$\alpha$, H$\beta$, and H$\gamma$
disk-center line profiles for
$\mathcal{F}_1=0$ (dotted line, \zmena{black}),
$\mathcal{F}_1=4\times 10^{11}$ (thick solid line, \zmena{blue}),
$6\times 10^{11}$ (thick dash-dotted line, \zmena{green}),
$8\times 10^{11}$ (thin solid line, \zmena{red}), and
$1.0\times 10^{12}$ (thin dash-dotted line, \zmena{violet}) $\rm{erg\,cm^{-2}\,s^{-1}}$.
No return-current effects are taken into account. Note that nonthermal
profiles almost overlap in this range of beam fluxes.
}
\label{fig:pr_bm}
\end{figure}

\section{Electron beam and return-current propagation}
\label{sec:beam}

We assume an electron beam that is accelerated in a coronal reconnection site
and injected into the cold chromosphere along the magnetic field lines. During
its propagation, the beam evolves under the influence of several processes
\citep{karlicky97}: (a) the beam generates the return current that decelerates
the beam in the return-current electric field, (b) the beam generates the
plasma waves causing the quasi-linear relaxation of the beam, and (c) the beam
electrons are decelerated and scattered due to collisions with the background
plasma particles. In the following model, we neglect the plasma wave processes
and the return current is taken in the form of runaway electrons
\citep{rowland85,vandenoord90,norman78}.
In this form, the
return-current losses are strongly reduced \citep{rowland85,karlicky04}.
Thus, only collisional losses, as described  by \citet{emslie78},
decelerate the electron beam in our case.

Let $\Phi=n_\BM v_\BM$ be the particle flux of the monoenergetic  beam of the
energy $E_\BM=m_\E v_\BM^2/2$, where $n_\BM$ is the density of the beam
electrons, $v_{\rm B}$ their velocity, and $m_\E$ the mass of the electron. According
to \citet{norman78} and \citet{karlicky04}, a fraction of background electrons
$\alpha=n_\RC/n_\E$ forms the current that moves in the opposite direction in
order to neutralize the electric current $e\Phi$ associated with the primary
beam. We use $n_\RC$ for the number density of the return-current electrons and
$n_\E$ for the number density of the background electrons. The neutralization
condition can be expressed as
\be en_\RC v_\RC=e\Phi\,.
\label{eq:neutral}
\ee
We assume that all the return-current electrons move with
the same superthermal velocity $v_\RC$.
\zmena{Note that the return-current flux only depends on the total flux of the
beam. For a realistic power-law distribution of the beam, the main part of the
beam flux is given by electrons with energy close to the low-energy cutoff of this
distribution. For these reasons and for simplicity, we consider
the monoenergetic beam in our model.
}

\begin{figure}[!ht]
\centering
\includegraphics[width=\columnwidth]{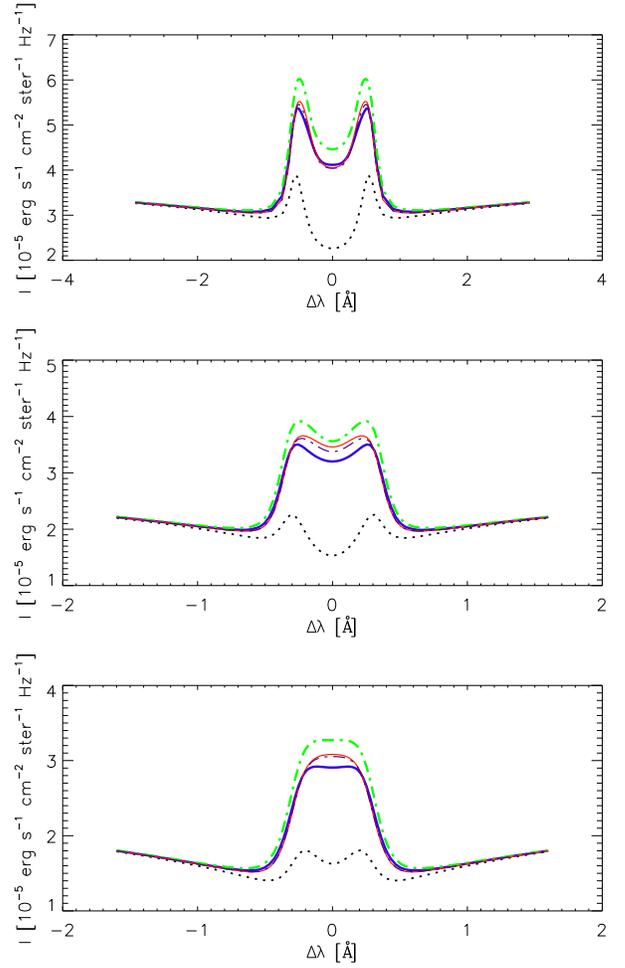}
\caption{
Same as Fig.~\ref{fig:pr_bm} plus including collisions with return-current 
electrons. The value of $\alpha$ is given by Eq.~(\ref{eq:ns}).
}
\label{fig:pr_ns}
\end{figure}

\begin{figure}[!ht]
\centering
\includegraphics[width=\columnwidth]{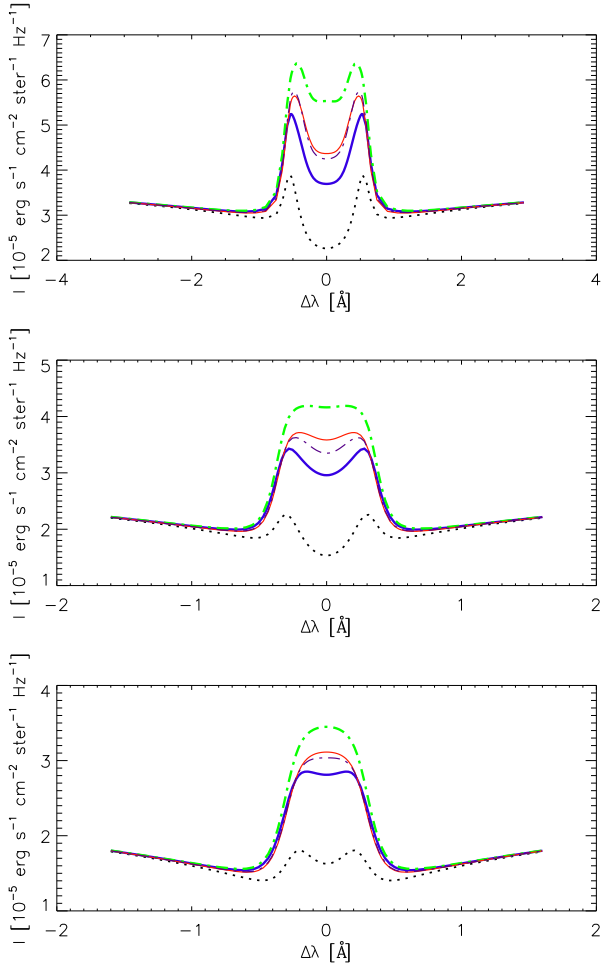}
\caption{
Same as Fig.~\ref{fig:pr_bm} plus including collisions with return-current 
electrons. The value of $\alpha$ is constant along whole beam trajectory.
}
\label{fig:pr_rc}
\end{figure}

We used two models for estimating $n_\RC$. First, following
\citet{norman78}, the number of runaway electrons can be estimated as
\be
\alpha=\frac{n_\RC}{n_\E}=\frac 12 \exp\left[ -\frac 12 \left(
\sqrt{\frac{\mathcal{E}_{\rm D}}{\mathcal{E}}}-
\frac{\mathcal{E}}{\mathcal{E}_{\rm D}} \right)^2 \right]\,,
\label{eq:ns}
\ee
where $\mathcal{E}/\mathcal{E}_{\rm D}=n_\BM v_\BM/n_\E v_{T_\E}$
\citep[see][for definition
of the electric field $\mathcal{E}$ generated by the electron beam
and the Dreicer electric field $\mathcal{E}_{\rm D}$]{karlicky04,xu05a},
and $v_{T_\E}$ stands for the thermal velocity of the background electrons. In
the second model we assume that the return current is formed by a fixed
relative number of background electrons everywhere in the upper chromosphere:
\be \alpha=\frac{n_\RC}{n_\E}={\rm const.}
\label{eq:alpha}
\ee
In this model, $\alpha$ is obtained
by averaging the values from the previous model over the Balmer line
formation layers. The resulting value for each beam flux
can be found in Table~\ref{tab:alphas}.

The normalized energy distribution of electrons can be locally expressed in
the form
\be
f(E)\,\D E=[c_{\rm M} f_{\rm M}(E)+c_\BM\delta(E-E_\BM)+
c_\RC\delta(E-E_\RC)]\,\D E\,,
\label{eq:eldistr}
\ee where $E_\RC=m_\E v_\RC^2/2$ is the return-current energy and $c_i$ stands
for the normalization coefficients (i.e., $c_\RC=n_\RC/(n_{\rm M}+n_\BM+n_\RC)$,
etc.). The index M stands for the background electrons (without the runaway
ones), and they obey the Maxwell-Boltzmann energy distribution.

\section{Hydrogen-electron collisions}

To take the effect of the beam/return current into account, we have to
calculate all the electron-hydrogen excitation rates,
the rates of ionization by electron impacts, and the rates of the
inverse processes. We use the
data for total collisional cross-sections for the bound-bound and bound-free
transitions by \citet{janev93}, retrieved through the
GENIE database (\texttt{http://www-amdis.iaea.org/GENIE/}).
We do not consider any atomic polarization or angular dependence of the
collisional processes in this work.
The excitation or deexcitation rate of the $n\to m$ transition
between the two shells can be calculated using the common formula
\be
C_{nm}=n_\E\int_0^\infty \D E\,\sqrt{\frac{2E}{m_\E}}f(E)\sigma_{nm}(E) \,,
\ee
where $\sigma_{nm}(E)$ is the total cross-section of $n\to m$
at impact energy $E$, and $f(E)$ is electron energy distribution
(\ref{eq:eldistr}). The cross-sections for deexcitation $m\to n$ ($m>n$)
were calculated using the $n\to m$ cross section and the
principle of detailed balance
\citep[e.g.,][]{jefferies68},
\be
\sigma_{mn}(E)=\frac{g_n}{g_m}\frac{E+E_{nm}}{E}\sigma_{nm}(E+E_{nm})\,,
\ee
where $E_{nm}$ is the excitation threshold of the transition and
$g_{n/m}$ the statistical weight of the given level.

The rates of the inverse process of ionization by an electron impact,
the three-body recombination ${\rm e+e+p\to\ion{H}{i}+e}$,
must be treated separately
due to the nonthermal nature of the problem. Using the arguments of detailed
balance \citep{fowler55,jefferies68} in the quasi-classical approximation of
the electron-hydrogen collisions, one can derive a formula for the recombination
rate.
Let us consider the ionization of level $n$ with the ionization energy
$E_n$ by an electron with energy $E_i$. Once cross-section
$S_{nc}(E_i,E_a)\equiv\partial\sigma_{nc}/\partial E_a$ of
the encounter after which we find the two electrons with energies $E_a$ and
$E_b=E_i-E_n-E_a$ is known, one can write the total recombination rate as
\be
C_{cn}=L\int_0^\infty \D E_a\int_0^\infty \D E_b\,
\frac{f(E_a)f(E_b)}{\sqrt{E_aE_b}}E_i S_{nc}(E_i,E_a)\,,
\label{eq:recomb}
\ee
where $L=n_\E^2n^2h^3/16\pi m_\E^2\approx 6.97\times 10^{-27}n_\E^2n^2$ in
cgs units, and $h$ stands for the Planck constant.
The expression (\ref{eq:recomb}) can be reduced to the well-known expression
$C^{\rm Thermal}_{cn}=2.06\times 10^{-16}n^2T^{-3/2}n_\E\exp(E_n/k_{\rm B}T)
C^{\rm Thermal}_{nc}$
in the particular case of Maxwell-Boltzmann distribution, where
$T_\E$ is the temperature of electrons and $k_{\rm B}$ the Boltzmann constant.
For the cross section $S_{nc}$ we used the approximate
data of \citet{omidvar65} at lower energies and the classical formula of
\citet{thomson12} at high energies.
In our case, the electron energy distribution function given
by (\ref{eq:eldistr}) leads to 9 terms 
that contribute to the recombination rates.
We verified numerically that, in all our models,
the rates of three-body recombination involving the nonthermal electrons
of the return current are more than one order of magnitude below the rates
of the same processes involving the thermal electrons.
Our tests show that neglecting the nonthermal three-body recombination
affects the resulting profiles in a very negligible way.
The three-body recombinations involving the beam electrons are completely
negligible since their rates are several orders of magnitude below
the thermal ones. Thus, we took only
the thermal term containing $m^2f_{\rm M}(E_a)f_{\rm M}(E_b)$ into account.

\section{Results}

\zmena{We calculated the NLTE radiative transfer for a 5-level plus continuum
hydrogen using the semiempirical 1D plane-parallel flare model F1 \citep{machado80} 
in which the
temperature structure was kept fixed;
in this way, we found the differential effects on the
H$\alpha$, H$\beta$, and H$\gamma$ lines \citep[cf.][]{kasparova02}.
We used the preconditioned equations of statistical equilibrium \citep{rybicki91}
and solved the coupled system
of NLTE equations by the accelerated lambda iteration (ALI) method.
For further details, see \citet{heinzel95}.
We found the equilibrium state
for several beam fluxes with or without the return current.}
The initial beam fluxes chosen in our calculations were:
$\mathcal{F}_1=4\times 10^{11},\,6\times 10^{11},\,
8\times 10^{11}$, and $1\times 10^{12}\,\rm{erg\,cm^{-2}\,s^{-1}}$.
If the fluxes were lower, the number of runaway electrons would decrease fast
and their energy would exceed the beam energy in most depths.
This gives a limit for using of this simple model.
Higher fluxes, on the other hand, would be unrealistic.
The initial energy of beam electrons was set to
$E_0=10$ keV.

\zmena{In Fig.~\ref{fig:pr_bm}, there are first three Balmer line profiles}
that result from the nonthermal bombardment by the primary beam. The effects
of return current were completely ignored. In this sense, these calculations
are similar to the ones of \citet{fang93} and \citet{kasparova02}.
In spite of their probably limited physical relevance, these profiles
are useful for demonstrating the effects of return currents in the
more appropriate models that follow.
Figure~\ref{fig:pr_ns} shows the situation where $\alpha$ is calculated
using Eq.~(\ref{eq:ns}); i.e., the relative number of runaway
return-current electrons is calculated at each depth in the atmosphere.
Finally, in Fig.~\ref{fig:pr_rc}, there are profiles for
the model with $\alpha$ constant along the atmosphere.
In the layers of Balmer line formation, $\alpha$ remains approximately
constant and its mean values are shown in Table~\ref{tab:alphas}.
Comparing Fig.~\ref{fig:pr_bm} with Figs.~\ref{fig:pr_ns} and
\ref{fig:pr_rc}, one can see that the effect of return current
is very significant: All three lines show a prominent increase emission
in the line center.

\begin{table}
\caption{The properties of the return currents. $\mathcal{F}_1$ stands for
the initial flux of the 10~keV beam, $\alpha$ is the mean relative number
of runaway return-current electrons in the Balmer lines formation region,
and $E_\RC$ stands for the typical energy of the return-current electrons in
these layers. Both $\alpha$ and $E_\RC$ are average quantities which
can roughly characterize the return-current properties in the region
of interest.
}
\label{tab:alphas}
\centering
\begin{tabular}{ccc}
\hline
$\mathcal{F}_1 [10^{11}\,{\rm erg\,cm^{-2}\,s^{-1}}]$ & $\alpha$ & $E_\RC [{\rm eV}]$\\
\hline
   4  & 0.01 & 100 \\
   6  & 0.05 & 14  \\
   8  & 0.11 & 6   \\
   10 & 0.16 & 3.8 \\
\hline
\end{tabular}
\end{table}

In the region of Balmer line formation, the energy of the return current can
be close to the excitation threshold of $n\ge 3$ levels of hydrogen,
and it remains approximately constant along an extended trajectory.
The beam is finally
stopped on a very short path. The overall path of the beam is, however,
sensitive to the initial energy of the beam.
In order to model the beam propagation and line formation accurately,
one has to interpolate the original F1 model of \citet{machado80} by a number
of grid points in the layers of the Balmer line formation.
Since the return-current energy and density are sensitive to the beam flux
(see Table~\ref{tab:alphas}), the resulting variation of the nonthermal
collisional rates leads to a significant variation in line profiles.
For both models under consideration, a maximum emission is found for
the beam flux of $6\times 10^{11}\,{\rm erg\,cm^{-2}\,s^{-1}}$, although the
resulting profiles from these models differ slightly from each other.
In contrast to the case of $4\times 10^{11}\,{\rm erg\,cm^{-2}\,s^{-1}}$, for which
we found the least emission among the studied flux intervals,
the return-current density is higher by a factor of 5. It leads to
a significant increase in nonthermal excitation rates.
The disagreement of the profiles at fluxes below $6\times 10^{11}\,{\rm erg\,cm^{-2}\,s^{-1}}$
is the result of the significant dependency of $\alpha$ on the beam flux and
atmospheric depth.
The values of $\alpha$ at low fluxes (shown in the Table~\ref{tab:alphas})
are only a rough approximation for the Balmer line formation layers, and
the model with $\alpha={\rm const.}$ seems to be less accurate 
than the one given by Eq.~(\ref{eq:ns}).
On the other hand, good correspondence between the models is found for high
fluxes. In this case, the variation in $\alpha$ is less sensitive to the
beam flux and
does not strongly vary with atmospheric depth.
Then, the $\alpha={\rm const.}$ model seems to give the appropriate results.
The reason the higher beam fluxes lead to a lower emission in the
lines is that the energy of the return current is not sufficient to excite
hydrogen atoms as can be seen in Table~\ref{tab:alphas}.

\section{Conclusions and outlook}
\label{sec:conclusions}

In this paper we used a simple model of the 10~keV electron beam
propagating in the chromosphere. We used two different models
of the return-current formation and calculated the differential effect
on the profiles of the hydrogen H$\alpha$, H$\beta$, and H$\gamma$ lines
of the semiempirical F1 model.

\zmena{
The return-current flux only depends on the total flux of the beam. For a
realistic power-law distribution of the beam, the main part of the beam
flux is given by electrons with energy close to the low-energy cutoff of
this distribution. Therefore, for simplicity we used the monoenergetic
beam in our model. Moreover, the excitation and ionization cross-sections
of low-energy return-current electrons are larger than those for beam
electrons, which makes the return-current effects on line core formation
stronger. Taking high-energy beam electrons into account (i.e. using
power-law distribution) would lead to increased emission in the line
wings due to penetration of those electrons into the deeper atmospheric
layers. However, the total flux of the beam in these layers is significantly
lower than the initial beam flux, and the return current and
corresponding effects are also strongly reduced.
}

Even our simple model shows that the effect of return current is very
important for future study of the hydrogen lines formation
since the energy of the return current can be expected to be on the order of
the excitation threshold energies of upper hydrogen levels, for which
the excitation cross-sections are high. Moreover, the fluxes $\Phi$ are high
enough to excite a sufficient number of atoms. As shown by \citet{karlicky04},
the collisional rates from the nonthermal collisions can dominate
the collisional rates in the Balmer line formation regions.
The two models used in this
work lead to similar results for higher energy fluxes, but the result differs
for lower fluxes. The excitation threshold effects seem to play an
important role for higher fluxes, but they are very likely only a consequence of the
monoenergetic model we used.
The difference between the two models shows the used approximations to be
incompatible for lower fluxes, where the the approximation of constant
$\alpha$ is not applicable.
\zmena{A detailed description of the energy distribution of the
return-current electrons would lead to more realistic line intensities.
This complex issue will be subject of a forthcoming paper,
which will also study the impact polarization of the Balmer lines.}

\begin{acknowledgements}
This work was partially supported by the grant
205/06/P135 of the Grant Agency of the Czech Republic,
partially by the grant IAA300030701 of the Grant Agency of the Academy
of Sciences of the Czech Republic, and partially by the project LC06014
Center for Theoretical Astrophysics.
\end{acknowledgements}

\bibliographystyle{aa}
\bibliography{bibs}

\begin{thebibliography}{21}
\expandafter\ifx\csname natexlab\endcsname\relax\def\natexlab#1{#1}\fi

\bibitem[{{Canfield} {et~al.}(1984){Canfield}, {Gunkler}, \&
  {Ricchiazzi}}]{canfield84}
{Canfield}, R.~C., {Gunkler}, T.~A., \& {Ricchiazzi}, P.~J. 1984, \apj, 282,
  296

\bibitem[{{Emslie}(1978)}]{emslie78}
{Emslie}, A.~G. 1978, \apj, 224, 241

\bibitem[{{Fang} {et~al.}(1993){Fang}, {Henoux}, \& {Gan}}]{fang93}
{Fang}, C., {Henoux}, J.~C., \& {Gan}, W.~Q. 1993, \aap, 274, 917

\bibitem[{{Fowler}(1955)}]{fowler55}
{Fowler}, R.~H. 1955, {Statistical mechanics} (Cambridge University Press)

\bibitem[{{Hawley} \& {Fisher}(1994)}]{hawley94}
{Hawley}, S.~L. \& {Fisher}, G.~H. 1994, \apj, 426, 387

\bibitem[{{Heinzel}(1995)}]{heinzel95}
{Heinzel}, P. 1995, \aap, 299, 563

\bibitem[{{Janev} \& {Smith}(1993)}]{janev93}
{Janev}, R.~K. \& {Smith}, J.~J. 1993, in Atomic and plasma-material
  interaction data for fusion, Vol.~4, 1--180

\bibitem[{{Jefferies}(1968)}]{jefferies68}
{Jefferies}, J.~T. 1968, {Spectral line formation} (A Blaisdell Book in the
  Pure and Applied Sciences, Waltham, Mass.: Blaisdell, 1968)

\bibitem[{{Karlick\'y}(1997)}]{karlicky97}
{Karlick\'y}, M. 1997, Space Sci. Rev., 81, 143

\bibitem[{{Karlick{\'y}} \& {H{\'e}noux}(2002)}]{karlicky02}
{Karlick{\'y}}, M. \& {H{\'e}noux}, J.-C. 2002, \aap, 383, 713

\bibitem[{{Karlick{\'y}} {et~al.}(2004){Karlick{\'y}}, {Ka{\v s}parov{\'a}}, \&
  {Heinzel}}]{karlicky04}
{Karlick{\'y}}, M., {Ka{\v s}parov{\'a}}, J., \& {Heinzel}, P. 2004, \aap, 416,
  L13

\bibitem[{{Ka{\v s}parov{\'a}} \& {Heinzel}(2002)}]{kasparova02}
{Ka{\v s}parov{\'a}}, J. \& {Heinzel}, P. 2002, \aap, 382, 688

\bibitem[{{Machado} {et~al.}(1980){Machado}, {Avrett}, {Vernazza}, \&
  {Noyes}}]{machado80}
{Machado}, M.~E., {Avrett}, E.~H., {Vernazza}, J.~E., \& {Noyes}, R.~W. 1980,
  \apj, 242, 336

\bibitem[{{Norman} \& {Smith}(1978)}]{norman78}
{Norman}, C.~A. \& {Smith}, R.~A. 1978, \aap, 68, 145

\bibitem[{{Omidvar}(1965)}]{omidvar65}
{Omidvar}, O. 1965, Phys. Rev., 140, A26

\bibitem[{{Rowland} \& {Vlahos}(1985)}]{rowland85}
{Rowland}, H.~L. \& {Vlahos}, L. 1985, \aap, 142, 219

\bibitem[{{Rybicki} \& {Hummer}(1991)}]{rybicki91}
{Rybicki}, G.~B. \& {Hummer}, D.~G. 1991, \aap, 245, 171

\bibitem[{{\v{S}t\v{e}p\'an} {et~al.}(2007){\v{S}t\v{e}p\'an}, {Heinzel}, \&
  {Sahal-Br\'echot}}]{stepan07protons}
{\v{S}t\v{e}p\'an}, J., {Heinzel}, P., \& {Sahal-Br\'echot}, S. 2007, \aap,
  465, 621

\bibitem[{{Thomson}(1912)}]{thomson12}
{Thomson}, J.~J. 1912, Phil. Mag., 23, 449

\bibitem[{{van den Oord}(1990)}]{vandenoord90}
{van den Oord}, G.~H.~J. 1990, \aap, 234, 496

\bibitem[{{Xu} {et~al.}(2005){Xu}, {H{\'e}noux}, {Chambe}, {Karlick{\'y}}, \&
  {Fang}}]{xu05a}
{Xu}, Z., {H{\'e}noux}, J.-C., {Chambe}, G., {Karlick{\'y}}, M., \& {Fang}, C.
  2005, \apj, 631, 618

\end{thebibliography}

\end{document}